\documentclass[pre]{revtex4}
\usepackage{epsfig}
\normalsize

\input{epsf}

\usepackage{graphicx}

\usepackage{amsbsy}

\usepackage{amssymb}

\usepackage{float}

\begin{document}

\title{ Agglomeration/aggregation and chaotic behavior in $d$--dimensional
spatio--temporal matter rearrangements. Number--theoretic aspects. }
\author{Adam Gadomski\dag and Marcel Ausloos\ddag}
\affiliation{\dag U.T.A. Bydgoszcz, Institute of Maths \& Physics,
Bydgoszcz PL--85796,  Poland \texttt{agad@atr.bydgoszcz.pl}}
\affiliation{\ddag  University of Li$\grave e$ge, SUPRATECS, Li$\grave e$ge B--4000, Euroland
\texttt{marcel.ausloos@ulg.ac.be}}
%
%

\begin{abstract}
Matter gets organized at several levels of structural rearrangements. 
At a mesoscopic level one can
  distinguish between two types of rearrangements, conforming to 
different close-packing or densification
   conditions, appearing during different
  evolution stages. The cluster formations appear to be temperature- 
and space-dimension dependent.
  They  suffer  a type of Verhulst-like saturation (frustration) when 
one couples the growing (instability)
  and mechanical stress relaxation modes together. They manifest a 
chaotic behavior both in space and time domains.
We pretend to offer a comprehensive and realistic
  picture of a material or mega-cluster formation in $d$ dimension.
\end{abstract}

\maketitle

\section{Inroduction}
\label{sec:1}
Matter organisations  at a mesoscopic (molecular--cluster) level 
typically manifest a multitude of
microstructural rearrangements. Cluster--cluster aggregations of proteins and/or colloids, phase 
separations, flocculation--coagulation
phase transformations, sol-gel systems, (wet) sand or rice piles, 
etc.,  are manifestations of
loosely-packed rearrangements, typically occurring under moderate or 
high temperature conditions.
In contrast ripened polycrystals, sintered powders, soap
froths and bubbles, and other cellular systems, constitute a type
of rearrangement that usually emerges in a (relatively) low temperature
limit and under certain  (''field dependent'') matter close--packing 
constraints.
Beside such an agglomeration, fracture, desaggregation, desorption, 
dissolution, and alike, can be thought to be
the ''inverse process'', finding its place in the opposite part of
the relevant phase diagram \cite{ma2}.

In all of them spatial as well as temporal signatures of chaotic 
behavior, due to matter reorganisations,
can be detected: They are temperature and space-dimension dependent. 
In particular, one can show rigorously that in the limit of
the spatial dimension going to infinity
loosely-packed agglomerations  become non--chaotic by suppressing 
totally their instability growing mode
since it is related to the nonequilibrium agglomerate's entropy 
\cite{evans}, while their closed-packed counterparts are not.
When the growing mode is coupled to a mechanical stress relaxation 
mode as a power law   via some
phenomenological relation of Hall--Petch--Griffith (H-P-G) \cite{HP} 
type (an Onsager-type conjecture \cite{lars}
of the present study), certain marks of Bethe--lattice frustration, 
related to a spatial overcrowding of the
Cayley--tree branches, appear in the (mean-field) approach - a kind 
of frustration qualitatively of very similar
type than that observed in Verhulst-type systems in an adequate time 
and parametric zone.

In the specific cases discussed in this review, however, by 
increasing the space dimension, $d$,
we automatically induce some increase of the possible number of 
degrees of freedom in the system.
Thus, when taking into account the coupling of the late-time growing  and relaxation modes, say, in a fairly synchronized viz power-law way of H-P-G type, one obtains that even though the material's relaxation goes slower than in the case when such a coupling is proposed in an unsynchronized (Debye-relaxation involoving, i.e. rapid) way, one is, however, able to establish or restore  an apparent dynamic microstructural order within the system the nonequilibrium (chaotic) measures of which are proposed below. There is, unfortunately, no way of establishing such an order when the coupling fails the power-law type synchronization requirement \cite{HP,lars}. 

Thus, when an ample space amongst the clusters is recovered by the 
system at its mature growing stage,
we consider that the system  successfully  tries to avoid a chaotic 
matter organisation in space.
Note that temperature may markedly help in surmounting the 
activation-energy barrier of the agglomeration,
especially when it is raised appropriately, whence when not 
"damaging' a possibly smooth  evolution of the system.
Full success is, however, guaranteed when the limit of $d\to\infty$ is reached.
If there is no chance for recovering the ample space,
the late-time growing stage is realized in a moderately chaotic way.
The mechanical stress relaxation, in turn, enters a readily chaotic 
regime, since the
  (nonequilibrium) entropy of the system  diverges to plus infinity.
The overall scenario resembles, in general, a formation of large 
(fractal) colloid
aggregates that typically occurs  with and without temperature and/or 
space-dimension dependent gravity factor domination,
like as if imposing some limits to gelation of colloids \cite{weitz}.

The paper is arranged as follows. In Section 2, we define both the
closely-packed as well as loosely-packed agglomerations,
calling the latter the aggregation throughout. In Section 3, we list some
qualitative signatures of chaos in matter-agglomerating systems, and 
refer briefly to different definitions as well as meanings of chaos.
In Section 4, we present quantitative measures of chaos signatures in 
systems of interest,
whereas in Section 5 we unveil number-theoretic measures, featuring a 
chaotic spatio-temporal behavior of them.
In Section 6, according to some suggestions given in \cite{rysiu},
on which much of our report is based, in order to see which 
agglomerations behave orderly or non-chaotically,
  we explore the limit of $d\to\infty$, and arrive at a certain 
interesting (perhaps, surprising) conclusion,
  favoring aggregation of matter, or some structural loosely-packed, 
and typically high-temperature, matter rearrangements 
-- in contrast to those emerging under close-packing low-temperature 
conditions.
We close the paper by offering a concluding address in Section 7.

\section{Agglomeration vs. aggregation of matter - a model description}
\label{sec:2}
Following \cite{kaye}, throughout the present study, we wish to distinguish
between the notions of agglomeration and aggregation of matter. By 
the former we mean an assembly
  of grains or molecular clusters, kept together by relatively strong 
forces ({\it e.g.} ionic), so that there is no
  easy  possibility
of taking the clusters apart, or destroying them.
  For the latter, because of the appearance of weak bonds between 
clusters, such as
Van der Waals or hydrogen types, the possibility of cluster 
separation becomes an observed tendency of the matter rearrangement
due to their weak bonds.
For a schematic explanation of the difference between both matter 
arrangements, see Fig. 1.
%
%
\begin{figure}
\centering
\includegraphics[height=6cm,width=10.5cm]{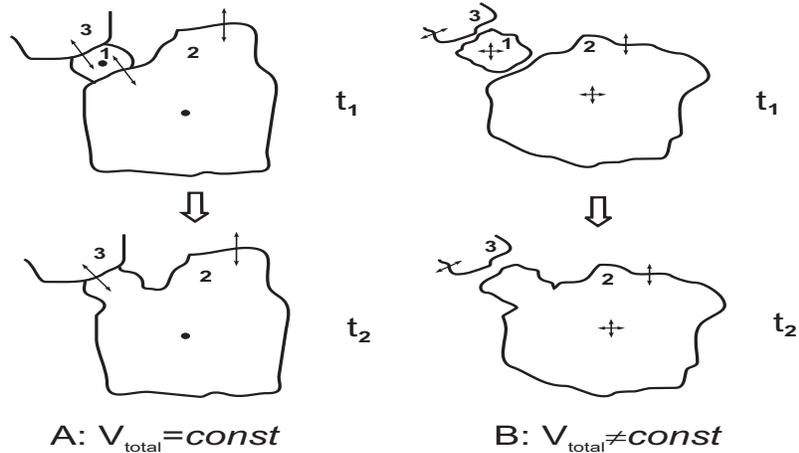}
%
%
\caption{Typical cluster-merging (three-grain) scenario for 
closely-packed (left, denoted by $\rm A$)
  and loosely-packed (right, denoted by $\rm B$) agglomerations. Two 
consecutive time steps $t_1$ and $t_2$ are shown.
  The former usually goes by a scenario with the preservation of the 
total agglomerate's volume
  (though in a more irregular way, when its logarithmic speed is 
measured, cf. Sections 3--6), whereas the latter does not 
\cite{physica}.
In the former, the clusters do not perform a translational motion but 
their boundaries may fluctuate
in time and space, even though they are quite strongly confined by 
their neighborhood. In the latter, an almost opposite situation in the 
  time-and-space domain is typically observed.
  Some void is left behind a loosely packed system}
\label{fig:1}       
\end{figure}

\subsection{Basic system of equations describing model matter agglomeration}
\label{s_sec:2-1}
As in previous work \cite{physica,chemphys} we begin with a local 
continuity equation
\begin{eqnarray}       \label{loccont}
{{{\partial }\over {{\partial t}}} f (v,t)} +
{{{\partial }\over {{\partial v}}} J (v,t)} = 0,
\end{eqnarray}
preferrentially supplemented by the corresponding initial (of 
delta-Dirac type as a first attempt) and boundary (typically, of 
absorbing type) conditions (IBCs).

In a few subsequent studies  
a ther\-mo\-dy\-na\-mic-kinetic description \cite{physica,vilar,chemphys} of model complex matter 
agglomeration has been worked out.
For the current \footnote{For a method to derive diffusion currents for different types of systems one is encouraged to look into \cite{vilar}} 
in the space of cluster volumes
\begin{eqnarray}
\label{JJ}
J (v,t) = - \Big[B(v) {\partial \over \partial v}
\Phi\Big] f(v,t) - {D (v)} {\partial \over \partial v}  f(v,t)  ,
\end{eqnarray}
has been used \cite{chemphys}, where $f(v,t)$ is the distribution of clusters of volume $v$: this 
means, that $f (v,t) dv$ is the (relative)
  number of clusters with size in the infinitesimal volume interval $[v,v+dv]$;
   $t$ is the time; $\Phi$ represents the physical potential, 
equivalent to the free energy of the system (see \cite{chemphys} for 
an explanation of the term).
   It is  assumed to be one of the  most relevant drivers of the 
agglomeration process at the mesoscopic level,  assuring its 
nonequilibrium character.
   In fact, the current (\ref{JJ}), in the form presented above, comes from a
   rigorous  derivation, starting from the Gibbs equation for the 
entropy production \cite{vilar}. It  has  quite strong foundations
    anchored in   nonequilibrium thermodynamics \cite{chemphys}. It is 
worth mentioning that upon inserting Eq. (\ref{JJ})
    into the continuity equation, Eq. (\ref{loccont}), one gets a 
second-order partial differential
     equation of the Fokker--Planck--Kolmogorov (F-P-K) type \cite{FPK}.

     For the dynamics  of such a system some routes to chaotic 
behavior have been sketched
     elsewhere by considering the (in)stability of Markov semigroups 
in \cite{pichor}.
      The mobility $B(v)$ is also defined in the configurational 
$v$--space and reads \cite{chemphys}
\begin{eqnarray}   \label{bv}
B(v) = {D\over {k_B T}} {v^\alpha }, \qquad \alpha = {{d-1}\over d},
\end{eqnarray}
where $D$ is a diffusion reference constant. Realize that the 
principal role of $D$ is to scale the
  time variable; $k_B$ represents the Boltzmann constant.
Note that the mobility $B(v)$ is related to the Onsager coefficient, 
$L(v)$, that appears in the derivation of the matter flux equation (\ref{JJ}) under a set of 
assumptions, mostly based
on the locality of the Kramers-type process \cite{vicsek,vilar}, etc.
$L(v)$, and hence $B(v)$, could be measured
by comparing the current and the thermodynamic force \cite{vilar}.
The quantity $D(v) = D v^\alpha $ is to be inferred from the 
Green--Kubo (G-K) formula \cite{dorfman},
so that there is some quite strong suggestion for deriving $B(v)$ 
both, experimentally as well as theoretically \cite{chemphys}.

There is a debate about a possible violation of the G-K formula 
\cite{dorfman,ivan}. For instance, it is proved
that for a gas of charged particles subjected to an external electric 
field, the mean mobility of a charged particle, based on the G-K 
formula,
is reliably well estimated for suitably small values of the external field.
Moreover, at a microscopic scale one observes a nonlinear (or 
chaotic) behavior of the particles,
which is, unfortunately, not reflected by the macroscopic 
(mobilility) measure. In our case, we  assume
algebraic correlations in $v$--space, for a phenomenological formula. 
The assumption seems to be as natural as possible: $D(v) \propto 
v^\alpha$;
that means that both the diffusivity $D(v)$ and the mobility $B(v)$ 
are proportional to the cluster hypersurface, ${s^{(d)}} := R^{d-1}$. 
It should be underlined that
it is, in our opinion, the common physical case in clustering phenomena,
and is working properly at the mesoscopic level considered in our approach.

Notice right here that exactly the same assumption has been used to 
model in a F-P-K way the formation of
surface nanostructure arrays \cite{bimberg}. Therein, an 
experimentally-observed passage between direct
curvature-dependent ripening of matter nano-islands (our 
densely-packed agglomeration), and  inverse ripening,
with an elastic-field caused contraction of growing quantum dots 
\cite{gczajk} (our sparsely-distributed agglomeration of matter),
  has been presented.

There are, however, matter agglomerations, for a given $T$, that do not 
conform usually to
\begin{eqnarray}   \label{bvs}
D(v) \propto B(v) \sim v^{\alpha} \simeq s^{(d)}.
\end{eqnarray}
To them belong both some physical-metallurgical transformations 
\cite{ma1}, such as martensitic,
and presumably, also certain phase orderings of non-diffusive kind, emerging in model biosystems, such as those occurring in 
lipid biomembranes \cite{peter}. Other than algebraic types of 
correlations in the hyperspace can likely be expected
  for these.
If a power law of the type given by Eq. (\ref{bvs}) can be kept for 
further modeling,
some additional correlations in time must complete a more 
comprehensive correlational proposal,
cf. \cite{graz}. Other types of correlations in the hyperspace, even 
if they   allow to get a general solution to the problem, may not 
accommodate
the boundary conditions \cite{chemphys}, so that one would expect 
either to be left with an unsolved specific problem or to encounter 
anomalous or
  irregular behavior of the agglomerating system 
\cite{physica,peter,graz}. In such a case another type of finite,
   instead of infinite boundary boundary conditions \cite{graz}, can 
sometimes give a remedy for the problem \cite{rysiu2}. Here, under the term  {\it infinite boundary conditions} \cite{physica} we typically understand the boundary conditions of absorbing 
(Dirchlet) type 
\begin{eqnarray}   \label{BCs}
f(v = 0, t) = f(v = {V_{clust}} ,t) = 0
\end{eqnarray}
in which the single cluster volume is taken at infinity, ${V_{clust}} = \infty$, whereas in case of the finite boundary conditions it assumes a finite value, $0 < {V_{clust}} < \infty$, cf. \cite{rysiu2}, and a discussion therein. Although the latter unquestionably seems more physical the former is more frequently used to reveal the evolutions in matter-agglomerating systems \cite{chemph} - this resembles to some extent a situation in statistical-thermodynamical systems undergoing an equilibrium phase transition: As such they are typically considered in the so-called thermodynamic limit (here, with a number of subunits going to $\infty$) under the mentioned agglomeration-oriented, e.g. condensation conditions, and the analogy would presumably extend over the examined nonequilibrium evolutions too \cite{rysiu2,chemphys}. 

\subsection{Thermodynamic potentials driving matter agglomerations}
In  previous work \cite{chemphys}
the analytic form of a (so-called) compaction potential was obtained, i.e.
\begin{eqnarray}   \label{philn}
\Phi(v) = \Phi_o ln{({R/R_o})},
\end{eqnarray}
where $\Phi_o$, $R_o$ - constants, and $R$ stands for some cluster radius.
Because
\begin{eqnarray}   \label{sim}
v\equiv {v^{(d)}} \sim R^d, \qquad d = 1,2,3, ... ,
\end{eqnarray}
one gets also $\Phi(v) \propto ln{({v/v_o})}$, where $v_o$ is a constant.

The  logarithmic potential assures the emergence of rather compact 
and curvature-involving structures,
whence the name of ''compaction potential''  \cite{chemph}.
It should be noted that $\Phi$ is an entropic potential \cite{chemphys}.
  Thus, it can be a cause of some desaggregation, or matter-influenced
  impingement effects, occurring within the overall aggregation space.

In a previous study on the phase transformation kinetics for loosely 
packed "diffusive"
agglomerates \cite{physica} we have written the matter flux of a 
purely diffusive nature
prescribed in configurational space as follows
\begin{eqnarray}
\label{JD}
J (v,t) = - {D (v)} {\partial \over \partial v}  f(v,t).
\end{eqnarray}
(The diffusion function  $D(v) = {D v^{\alpha }}$
is  proportional to the cluster (grain) surface.)

Both closely-packed and loosely-packed agglomerations follow from the 
general form (\ref{JJ}).
Indeed, the loosely-packed case is  obtained
  when the first (drift) term in r.h.s. of  (\ref{JJ}) can be neglected.
Formally, $B(v) \to 0$ when $T\to \infty$. From the physical point of 
view, it corresponds to sufficiently high temperatures  $T \ge 
T_{pass} > 0$,
where $T_{pass}$ can be treated as  a cross-over temperature
\footnote{Borrowing from the nomenclature of phase transitions and 
critical phenomena one might sometimes opt for calling it the 
threshold temperature} above which
  the agglomeration  takes place exclusively by yielding 
loosely-packed microstructures.
However, the drift term in r.h.s. of  (\ref{JJ}) depends both on $v$ and $T$.
It tends {\it uniformly}, which means independently of $v$, to zero at the
high temperature limit  if
\begin{eqnarray}   \label{cond1}
B(v){{\partial \Phi(v)}\over{\partial v}} \simeq {\rm C} = const.
\end{eqnarray}
Then for a given system, temperature $T_{pass}$ does not depend on $v$
and  looks consistently defined.

Some additional argumentation can be provided that such a constant 
(limit) ${\rm C}$ exists and is well-defined.
Namely, when applying both (\ref{bv}) and (\ref{sim}) one sees with 
sufficient accuracy that
\begin{eqnarray}   \label{cond1C_E}
{\rm C} \propto {1\over \Delta R} \times {{\Delta\Phi (v)}\over {k_B T}}.
\end{eqnarray}
This means that ${\rm C}$ is essentially determined by a product 
involving two contributions: a certain curvature-like term,
  $\kappa = 1/ \mid \Delta R \mid$, and some dimensionless energetic 
argument, ${\epsilon _E} = {{\mid \Delta\Phi (v)\mid}/ {k_B T}}$.
   The above claimed high-temperature limit, with the cross-over 
temperature $T_{pass}$ as a reference temperature characteristic of a
   system of interest, would naturally demand $0 < \epsilon _{E} <<1 $ 
while, because of approaching the mature growing stage any change
    in the cluster radius must be small, $0 < \mid \Delta R\mid << 1$, 
and therefore, its inverse would tend to some big value, i.e. 
$\kappa >> 1 $.
    Thus, ${\rm C}$ will take on a finite value.
It is believed that for certain agglomerations under readily 
high-temperature conditions
  it will eventually acquire a small value\footnote{Such a belief 
comes undoubtedly from the fact that we
  offer our approach for systems evolving in an overdamped regime, 
such as those of biopolymeric type. For them the Reynolds number
  is typically of the order of $10^{-3}$, i.e. very low, so that the 
mobility per se, even for a single biomolecule but also for a
   molecular cluster, must clearly be of negligible value 
\cite{vicsek}, regardless of whether we measure it in the $v$--space 
or, what is usually done,
   in a position space}, that means, $0 < {\rm C} << 1$ naturally holds.
It is a case when the potential
\begin{eqnarray}   \label{phiv}
\Phi(v) \propto v^{1 - \alpha } \simeq v^{1/d}.
\end{eqnarray}

In \cite{bimberg} a condition of setting the current equal to zero, $J = 0$,
  has been chosen to balance diffusive and non-diffusive terms in the F-P-K type
   description, cf. \cite{vilar}, aiming at getting a proper behavior of the metastable 
nanostructure \cite{gczajk} arrays.
   We are of the opinion that such a proposal is legitimate in the 
relatively low-temperature domain.
    When the temperature is raised, but  agglomeration is still 
allowed to occur, the proposal may fail.
    Thus, the above is a possible solution for the high-temperature limit.
    A type of localization of the Gaussian distribution, 
characteristic of the inverse ripening
    (a metastable state of the nanostructure evolution) can also be 
obtained within the present modeling, cf. \cite{physica}.
This is the case of Eq. (\ref{JD}) when in a (readily) mature growing 
stage, since the single volume $v$
of the cluster does not change much. As a matter of fact, there is no 
small-cluster population available
for merging (Fig. 1), i.e. $D(v) \to const.$, which nearly 
corresponds  to the high-temperature criteria of Eq. (\ref{cond1}), 
or equivalently  Eq. (\ref{cond1C_E}).
In so doing,  Eq. (\ref{JD}) represents the 1st Fick law in its 
standard form. Upon inserting it into
  Eq. (\ref{loccont}) one immediately arrives at the 2nd Fick law (in 
the configurational space) with its
  standard Gaussian solution, the metastable case being emphasized in 
\cite{bimberg}.

The above potential form (\ref{phiv}), designed for loosely-packed 
agglomeration, seems legitimate here:
Note that the 'force' ${F_{c-c}} \propto {{\partial \Phi(v)}/ 
{\partial v}}$ behaves like

\begin{eqnarray}   \label{fgg}
{F_{c-c}} \propto {1\over v^\alpha} \simeq {1\over s^{(d)}},
\end{eqnarray}
because  $v\sim R^d$. Thus, $F_{c-c}$ acts as the inverse of the area 
of the cluster hypersurface, $s^{(d)}$,
  which implies that the smaller the area is, the bigger the force 
acting on the cluster can be, this way
  impeding the formation of new clusters, which would contribute to an 
aggregate's density increase.
Qualitatively, a similar dependence is found for the closely-packed 
matter agglomeration:
from (\ref{philn}) one gets, as above, for the 'force'
\begin{eqnarray}   \label{ffgg}
{F_{c-c}}  \simeq {1\over v^{(d)}}.
\end{eqnarray}
Here, $F_{c-c}$ acts as the inverse of the hypervolume of the 
cluster, $v^{(d)} := R^d$, which makes
  a clear difference between closely-packed and loosely-packed 
agglomerations, presumably leading
  to a certain  relaxation of the surface tension conditions for 
loosely-packed clusters-containing systems \cite{chemph}.

Referring further to (\ref{cond1}) and using
the similarity relation, Eq. (\ref{sim}), one gets
\begin{eqnarray}   \label{phiR}
\Phi(R) \propto  {\Phi({R_o})} {R\over R_o},
\end{eqnarray}
where
\begin{eqnarray}   \label{phiRo}
{\Phi({R_o})} = {k_B T\over {D_{\alpha }}} R_o, \qquad T \ge
  T_{pass} ,
\end{eqnarray}
and consequently, $\Phi(R) \propto R$.
Moreover,
\begin{eqnarray}   \label{Dalfa}
{D_{\alpha }} = D {(1 - \alpha )}.
\end{eqnarray}
  $R_o$ can now be specified  to be the initial cluster radius.  Note 
that $D_{\alpha }$ is a $d$--dependent quantity.

\subsection{Cluster volume fluctuations as reliable characteristics 
of matter agglomeration}
Aggregations and agglomerations  emerge in a fluctuating changing 
medium. Therefore, any
reasonable quantitative attempt on resolving the fluctuation impact 
on their speed is worth
examining here. In what follows,  let us propose an evaluation of the 
reduced variance
\begin{eqnarray}       \label{frf}
{{\sigma }^2 (t)} = {{<v^2 (t)> - {<v^1 (t)>}^2}\over {{<v^1 
(t)>}^2}}\equiv  {{<v^2 (t)>\over {<v^1 (t)>}^2} - 1} ,
\end{eqnarray}
as a direct measure of the cluster volume fluctuations.

The notation used in Eq. (\ref{frf}) refers to the statistical moments
\begin{eqnarray}       \label{mom}
{<v^n (t)>} = \int_0^\infty v^n f(v,t) dv \qquad  n = 0,1,2, ...
\end{eqnarray}
of the stochastic process, where the matter agglomeration is usually 
described by the  local
  continuity equation, Eq. (\ref{loccont}).

The explicit solutions, $f(v,t)$--s, have been presented elsewhere 
\cite{physica,chemphys,graz}, and refs. therein.
The zeroth moment, $<v^0 (t)>$, is related to the average number of 
molecular clusters in the system,
  and usually  shows  an algebraic decrease with time \cite{physica}. 
The first moment, $<v^1 (t)>$, is related to
  the total volume
which is a constant value for closely-packed agglomerations 
\cite{chemph} and an increasing function of time
for loosely-packed agglomerations \cite{physica}, cf. Fig. 1. From 
the expressions of  both moments, it follows
that the average cluster radius, $R_{av} (t)$, behaves as a power law 
in  time, with a
growth exponent $1/(d+1)$ that apparently contains some signature of 
random close-packing of matter
by having included the super-dimension $d+1$ \cite{chemphys,zallen}. 
($d+1$ tells us something about the minimum
number of non-overlapping neighbors of a given cluster in a 
$d$--dimensional space.) These constitute the main characteristics
of the model agglomeration/aggregation  process in its late-stage 
($t>>1$) limit.

The question remains about asymptotic values of the moments $<v^n 
(t)>$ that must be known when applying
  formula (\ref{frf}). For closely-packed agglomerations, the moments 
are found to obey a power law \cite{chemph}
\begin{eqnarray}       \label{frfCP}
{<v^n (t)>} \sim t^{(n-1)/(2-\alpha )} \qquad (n=0,1,2), \quad t >> 1 ,
\end{eqnarray}
whereas for   matter aggregation  one  finds another power law \cite{physica}
\begin{eqnarray}       \label{frfLP}
{<v^n (t)> } \sim t^{[(n-1)+\alpha ]/(2-\alpha )} \qquad (n=0,1,2), 
\quad t >> 1.
\end{eqnarray}
Notice, that for $\alpha = 0$ ($d=1$) both power laws above approach 
the same form, namely ${<v^n (t)>} \sim t^{(n-1)/2}$.
When utilizing (\ref{frf}) and (\ref{frfCP}) it appears that for 
closely-packed agglomerations,
  ${{\sigma }^2 (t)}$ can be fully identified with the inverse of 
$<v^0 (t)>$ (the average number of clusters),
  {\it cf.} \cite{chemph} for details, what because of the constancy 
of $<v^1 (t)>$, leads to ${{\sigma }^2 (t)}
   \propto V_{sp} (t)$, where ${V_{sp}}\equiv V_{sp} (t) \simeq <v^1 
(t)>/<v^0 (t)>$, and can be termed the mean
   specific volume of the tightly-packed agglomerate, being 
equivalent to the inverse of its mean number density.
   The specific volume fluctuations read
\begin{eqnarray}       \label{frfv}
{{\sigma }^2 (t)} \propto {t^{d/(d+1)}},
\end{eqnarray}
and if $d\to\infty$, ${{\sigma }^2 (t)} \simeq V_{sp} \propto t$.

When using (\ref{frf}) and (\ref{frfLP}), however, it turns out that 
for loosely-packed agglomerations ${{\sigma }^2 (t)}$
is a quantity equivalent to the average cluster radius $R_{av} (t)$, 
see \cite{physica,chemphys}.
They behave in time as
\begin{eqnarray}       \label{frfR}
{{\sigma }^2 (t)} \propto {t^{1/(d+1)}} .
\end{eqnarray}
When $d\to\infty$, ${{\sigma }^2 (t)} \simeq R_{av} (t) \to const$, 
which means, that on average the system ceases to grow.
Note that the standard diffusional regime, is always characterized by 
the one-half exponent, is achieved exclusively for the aggregations
  in $d=1$ because the only linear  characteristic is $R_{av}\equiv 
R_{av} (t)$: Note that $V_{sp} (t)$ is not a linear characteristic,
  since $V_{sp} (t) \propto {{[R_{av} (t)]}^3}$ usually holds. Here 
the $d=1$--case must clearly be disqualified as
   standard diffusional, cf. Eq. (\ref{frfv}).

Commenting on the last relations, (\ref{frfv}) and (\ref{frfR}),  one 
might furthermore conclude that they reflect a well-known
Onsager conjecture that the fluctuations in a system undergo the same 
type of changes as the corresponding macroscopic dynamic
variables \cite{lars}: Here one may think of the specific volume of 
the agglomerate and the grain radius, and their behavior in the 
late-time domain, respectively.

\subsection{Coupling the instability (growing) and mechanical stress 
relaxation modes of matter agglomeration}
\label{s_sec:2-2}
Poisson was likely the first who recognized that viscoelastic 
properties of fluids and
  solids can reasonably be compared in a suitable, mostly short-time 
domain, though
  the specification of the domain must   be more precise  for 
specifying the systems of interest.
  Maxwell successfully followed the ideas of his famous French 
predecessor, arriving at his well-known,
  in general non--Markovian, model of relaxation \cite{evans}. In what 
follows we present our Maxwell--model-based
   ideas on how to distinguish between the two agglomerations under 
study, and how to switch on a kind of
   coupling between the (late-stage) growing and relaxational modes in 
the viscoelastic $d$--dimensional
   matrix that we investigate. The existence of the coupling seems to 
be experimentally justified,
   see \cite{weitz,ivan,bimberg}, and involves generically the 
viscoelastic nature of the mega-cluster
    late-stage formation \cite{evans}.

Thus, the afore presented rationale toward quantifying the 
fluctuations of the system can be
strengthened with a supporting phenomenological argumentation. The 
idea comes from a "coupled"
diffusion-relaxation picture that appears in such a complex system. 
In any diffusion-migration
growing process, the mechanical strain-stress fields play such a role 
as well. In our case, such
a situation can be safely expected in the temperature domain $T\le 
T_{pass}$. Another
type of relaxation of the stress field, say $\sigma _m$, is expected 
to prevail when the closely-packed
  agglomeration conditions are met. A different behavior may be 
observed when the closely-packed agglomeration
   conditions are lost for the first time, that is,   at $T= 
T_{pass}$, when the loosely-packed context
   appears. In both temperature regimes, the relaxation of $\sigma _m 
(t)$ over the course of time,
   is very likely to go in a way
    essentially described by the current (\ref{JJ}). This is expected 
to occur \cite{chemph} presumably under
     (nearly) homogeneous strain conditions, $\epsilon _m \approx 
const$, for $t >> 1$.
For an additional motivation of coupling  matter agglomeration and 
stress relaxation
picture, related to fracture phenomena, see \cite{ma2,chemph}.

 From \cite{chemph} it can be learned, that in the absence of 
non-Arrhenius or fractal type kinetics,
  seemingly modifying the diffusion coefficient $D (v)$ \cite{graz}, 
one expects the Maxwell dashpot-and-spring model
  to reflect properly the relaxation behavior.
We wish to set up here a phenomenological picture, showing that both 
agglomeration and mechanical stress relaxation,
  where the stress relaxation takes place under slow growth 
conditions, proper of a mature growing stage in a viscoelastic
   multiphase medium \cite{vicsek,ivan}, are coupled processes 
\cite{weitz,bimberg}. To work out the problem quantitatively,
    we will represent one of the two contiguous and matter-exchanging 
clusters in the agglomerate, say cluster (grain) $1$,
     as an expanding one, equivalent to the spring, growing at the 
expense of its neighbor, to be named cluster (grain) $2$,
     i.e the dashpot, to which, according to the Maxwell model, the 
contracting action should be assigned,
{\it cf.}, Fig. 2 for details; see \cite{chemph}.
%
%
\begin{figure}
\centering
\includegraphics[height=6cm,width=10.5cm]{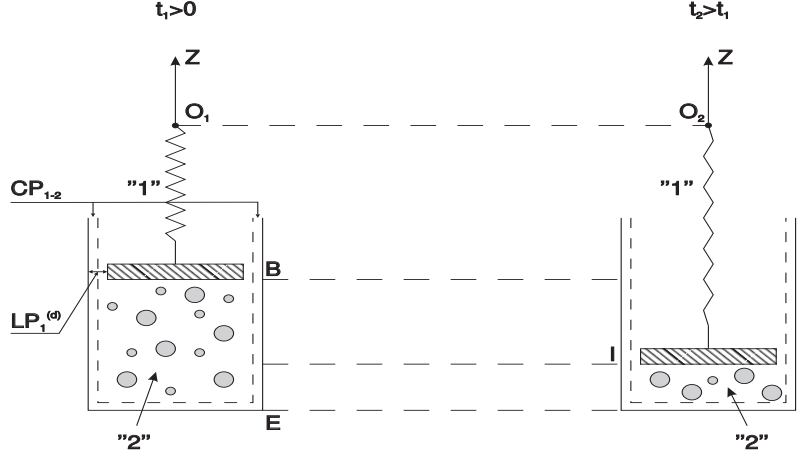}
%
%
\caption{Maxwell sequential spring-and-dashpot (quasi-fractional) 
model with narrow ($d$--independent) and wide ($d$--dependent)
gaps, shown schematically in two subsequent time instants $t_1$ and 
$t_2$, where $t_2 > t_1 > 0$,
from left to right, respectively. Grain "1" consists of the spring 
and the piston's upper wall, to
which the second end of the spring is attached, while its first end 
is mounted either on $0_1$ or
$0_2$, from left to right, respectively. Grain "2" consists of the 
viscous medium inside the cylinder as
well as the inner wall of the piston.
The cylinder's walls complete the overall model structure of the 
viscoelastic grains.
The material exchange between "1" and "2" is assured by the
existence of the gaps: narrow ${CP}_{1-2}$ gaps in case of 
closely-packed  agglomeration, and some two wider (here, represented
by the left-hand side gap, ${{LP}_1}^{(d)}$) in case of the aggregation.
Therefore, the piston-and-cylinder system, containing a viscous 
fluid, here composed of big and small particles, is either more 
(densely-packed agglomeration) or less (undensely packed formation) 
leakproof.
The overall material exchange
is caused by spring expansion along $z$ axis, which results here in a 
longitudinal expansion of grain "1"
at the expense of grain "2", {\it cf.} {\it 
http://www.j-npcs.org/abstracts/vol2000no4.html}.
Notice formally that: ${{0_1}E}={{0_2}E}$, and for $t_1$ one has $z_1 
(t_1 ) = {{0_1}E}$,
$z_2 (t_1 ) = {BE}$ as well as  for $t_2$ one gets $z_1 (t_2 ) = {{0_2}I}$,
$z_2 (t_2 ) = {IE}$, which results in grain expansion-contraction 
behavior, like
${{0_1}B}<{{0_2}I}$ and ${BE}>{IE}$, when mutually comparing the 
distances along  the $z$ axis at
$t_1$ and $t_2$, respectively}
\label{fig:2}       
\end{figure}

For the system with non-wide gaps, the Maxwell model conditions
  are almost satisfied, so that the two-cluster action can be extended over
  all pairs of contiguous clusters until the expanding (growing) 
eventually survive.
  In a next step, the same kind of competition appears as in the well 
argumented Laplace-Kelvin-Young
   scenario suitable for cellular systems \cite{chemphys}. This 
picture holds in the closely-packed context.

In the loosely-packed context (a system with wide gaps),  we may have 
qualitatively almost
the same picture \cite{evans} but with several  differences which 
implies that cluster
  expansion would not be  likely so vigorous. Since the corresponding 
gap is wider, therefore untight,
   the fluid leakage might be more pronounced. Thus, the fluid 
response against the piston wall is weaker,
   and the Maxwell type stress relaxation no longer applies, cf. the 
caption of Fig. 2.

The stress relaxation can   be described by introducing an exponent 
$\chi $ in the Maxwell-like
quasi-fractional model presented  here below.  This exponent should 
be, in general, $d$-dependent, and points to
a difference when comparing with the classical Maxwell model 
\cite{evans}. Here, we offer a
coupled   matter diffusion and stress relaxation picture, but for a 
random walk performed in the
configurational space \cite{chemph}.
As is known, the Maxwell stress relaxation picture leads to an 
exponential decay of the stress:
\begin{eqnarray}       \label{expM}
\sigma _m (t) \sim exp (-t/\tau _M ),
\end{eqnarray}
where $\tau _M $ is a reference time for the concentrated clusters 
\cite{evans} to be eventually inferred from
the Einstein-Stokes-like formula \cite{vicsek,chemph}.  This behavior 
holds for $T < T_{pass}$. As mentioned above,
  for $T\ge T_{pass}$ we propose
\begin{eqnarray}       \label{Max}
{{d\sigma _m}\over {dt}} + {{{\sigma _m}^{\chi }} \over {\tau  }} = 0,
\end{eqnarray}
where the above is usually true when the internal strain field, 
$\epsilon _m $ is practically constant,
see above. When solving (\ref{Max}), one obtains
\begin{eqnarray}       \label{Maxs}
{\sigma _m (t)} \sim {{(t/\tau  )}^{-1/(\chi -1)}}, \qquad t >> 1,
\end{eqnarray}
where $\chi = {2 d + 3}$; about $\tau$, see \cite{chemph} or \cite{vicsek}.
Notice that for $\chi = 1 $ and $\tau = \tau _M$ in Eq. (\ref{Max}) one gets
the solution (\ref{expM}); for $\chi \ne 1$ Eq. (\ref{Maxs}) is the 
only solution
to the relaxational problem as stated.
When comparing Eqs. (\ref{expM}) and (\ref{Maxs}) one sees that the
relaxational response goes slower for the late-time loosely packed 
aggregational context than for
  its densely-packed agglomerational counterpart.


%
%
%
\section{Qualitative signatures of chaos in matter--agglomerating system}
\label{sec:3}

Let us consider
a few qualitative signatures of chaos in matter--agglomerating 
systems from the literature.
Such certain signatures for systems  of the type studied in the present work
  are summarized in Table 1.

\begin{table}
\centering
\caption{Qualitative signatures of chaos in a model 
matter--agglomerating system of interest}
\label{tab:1}       
%
%
\begin{tabular}{lll}
\hline\noalign{\smallskip}
Item &                              Signature &                Refs. \\
\noalign{\smallskip}\hline\noalign{\smallskip}
One:   & Entropic system of molecular-chaotic behavior& 
\cite{dorfman,vicsek,schmelz} \\
Two:   & Lack of matter - depletion zones around the charged 
clusters& \cite{henk,vicsek,chemph} \\
Three: & Competition-and-loss effect: (un)tight spring-and-dashpot 
model& \cite{wloch,zallen,evans} \\
Four:  & IBCs of normal (e.g., Neumann) or abnormal type should be taken 
appropriately & \cite{rysiu2,rysiu,graz} \\
Five:  & Bethe-type frustration in coupled relaxation and 
late-growing events & \cite{ma2,zallen,wloch} \\
Six:   & Growth viz instability: random close-packing with its 
$d+1$--account & \cite{zallen,schmelz,chemphys} \\
Seven: & Nonequilibrium entropy measures viz mean-harmonic speeds  & 
\cite{dorfman,schuster,evans} \\
Eight: & Entropic potential(s) assuring nonequilibrium character of 
the phenomenon & \cite{dorfman,evans,chemphys} \\
Nine:  & G-K type construction of $D(v), B(v)$, and its consequences 
& \cite{dorfman,lutz,evans} \\
Ten: & Diffusion-space pre-chaotic (Fibonacci) feature by 
$D_{\alpha}$ & \cite{schuster,dorfman,physica} \\
\noalign{\smallskip}\hline
\end{tabular}
\end{table}
The items stated in Table 1 do not exclude other possible forms
  to chaos, or its signatures, in matter-agglomerating systems.
  We do not pretend to describe all of them, or even their majority.
   For routes to chaos recommended from physical point of view one would
   usefully consult \cite{lutz,schuster}; which routes, or scenarios 
of chaos, are recommended by mathematicians,
   especially when a partial-differential-equation formalism of F-P-K 
type is effective, can be found in \cite{rysiu},
    and in refs. therein.

\section{Some quantitative measures of chaos signatures in 
matter--agglomerating system}
\label{sec:4}
In \cite{chemph} some entropic-like nonequilibrium measures of growth
\begin{eqnarray}       \label{LPsp}
{{\nu _{sp}}^{(d)}} = {\Big({ln{[{\sigma }^2 (t)]}\over 
{ln(t)}}\Big)}_{{for}\quad t>>1} , \quad d = 1,2,3, ...
\end{eqnarray}
as well as for the mechanical stress relaxation evolution
\begin{eqnarray}       \label{LPrel}
{{\mu _{sp}}^{(d)}} = {\Big({- ln{[{\sigma _m} (t)]}\over 
{ln(t)}}\Big)}_{{for}\quad t>>1} , \quad d = 1,2,3, ...
\end{eqnarray}
   have been proposed.
   This seems to be working most appropriately in a 
growth-and-relaxation synchronization metastable regime
\begin{eqnarray}       \label{hp}
{\sigma _m} \sim {{R_{av}}^{-1/2}} \sim {\sigma ^{-1}} ,
\end{eqnarray}
which represents the H-P-G condition \cite{HP}
\begin{eqnarray}       \label{hp2}
{\sigma _m} \sim {{R_{av}}^{-1/2}}  ,
\end{eqnarray}
appropriate for the fluctuational late-time regime \cite{vicsek} of interest here.
  Bear in mind that if certain empirical modifications of the
formula (\ref{hp2}) are applied toward obtaining a specific form, 
interconnecting  Eq.
(\ref{LPsp}) with Eq. (\ref{LPrel}), one gets something like
\begin{eqnarray}       \label{hp3}
{\nu
_{sp}}^{(d)} = q {\mu _{sp}}^{(d)},
\end{eqnarray}
where typically $q > 2$. In the classical H-P-G limit ${\nu
_{sp}}^{(d)} = 2 {\mu _{sp}}^{(d)} $ holds.
However $q$ may also strive for obtaining superplastic effects, i.e.
when taking on fractional values, {\it cf.} \cite{chemphys}, and
refs. therein. This is sometimes termed in physical-metallurgical literature the inverse H-P-G effect.

Because of Eq. (\ref{frfR})
\begin{eqnarray}       \label{LPsp2}
{{\nu _{sp}}^{(d)}} = {1\over {d+1}}, \quad d = 1,2,3, ... .
\end{eqnarray}
Realize that formulae (\ref{hp}) and (\ref{hp2}) might again be 
interpreted in terms of the Onsager conjecture \cite{lars,evans}, see 
above.
Since the overall exponent in (\ref{Maxs}) reads
\begin{eqnarray}       \label{x>0}
{1\over {\chi -1}} = {1\over {2(d+1)}},
\end{eqnarray}
which is exactly one half of the growth exponent ${\nu _{sp}}^{(d)}$ 
given in (\ref{frfR}), see  (\ref{LPsp2}) too, one consequently 
provides
\begin{eqnarray}       \label{LPrel2}
{{\mu _{sp}}^{(d)}} = {1\over {2(d+1)}}, \quad d = 1,2,3, ...
\end{eqnarray}
Let us emphasize here that $1\over {\chi -1}$ stands for the 
so-called Nutting exponent for relaxation,
and can be interpreted in terms of the loss tangent, that means, a 
well-known dissipation factor in the
relaxation phenomena, mostly in dielectric (e.g., macromolecular) 
environments \cite{chemph}.
\section{Number-theoretic measures of spatial and temporal 
irregularities in aggregation-agglomerating systems}
\label{sec:5}
%
It is interesting to note here that $\chi = \chi (d)$ , i.e.
\begin{eqnarray}       \label{odd}
\chi (d) = 2(d+1)+1 .
\end{eqnarray}
A certain generator of the
Bethe--lattice elements, is recovered
starting from the 3-bond (initial) generator for $d=0$, and 
continuing with $d$,
upon identifying $d$
as the numbers of emerging bonds in a gelling system \cite{weitz}.
  This is a  very
  useful tool for the mean-field description of gels, and other multi 
bond-containing systems. T

In this way, an odd number Bethe--lattice generator for subsequent 
$d$--s can be offered, see Fig. 3.
%
%
\begin{figure}
\centering
\includegraphics[height=6cm,width=10.5cm]{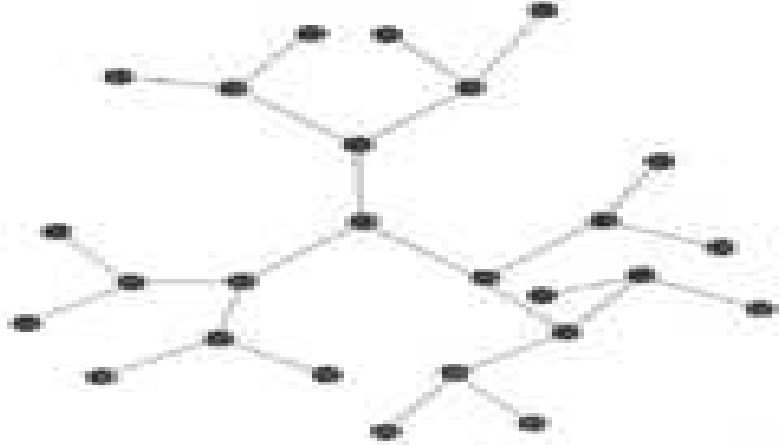}
%
%
\caption{An example of a not much developed Bethe lattice, which by 
itself manifests a frustration because
of "having problems" with containing all of its nodes in the 
available $d$--dimensional space \cite{wloch},
to some extent so as, for example, the population of Verhulst fellow 
countrymen does in the available Belgium territory \cite{pastijn}}
\label{fig:3}       
\end{figure}

Another, equally interesting observation can be offered, namely
\begin{eqnarray}       \label{HR}
{2\over {\nu _{sp}}^{(2)}} = {{1\over {\nu _{sp}}^{(1)}} + {1\over 
{\nu _{sp}}^{(3)}}}.
\end{eqnarray}

This means that for loosely-packed agglomerations the harmonic-mean 
rule for the growth
  speed is exactly fulfilled. {\it Mutatis mutandis}, one can expect 
the same type of rule, Eq. (\ref{HR}), for ${\mu _{sp}}^{(d)}$.
Let us recall that the fluctuations ${{\sigma }^2 (t)}$ have been proposed as
  a reliable criterion of differentiating between  aggregation and 
agglomeration,
  and that an efficiency (harmonic-rule, see (\ref{HR})) additional criterion, derived 
from the constructed
  fluctuational proposal, supports the aggregation in dimension $d$ 
($d=1,2,3$), with an emphasis placed
  on $d=2$, where 'golden-ratio-like' or harmonic-mean properties are 
in favor. The mean-harmonic speed
  implies that the center of mass of a moving body, referred consequently to
  as the molecular cluster,  may not span the same distance, say $s$, back and
  forth, during a time period. This leads to a  quite realistic 
quantification of a
  mean speed on the distance $2s$, and also shows that not an arithmetic mean of
  the back and forth speeds but a harmonic mean applies here. Such a 
schedule can
  likely be extended over the aggregation that essentially relies on
  random matter attachments and/or detachments of particles 
\cite{henk,physica}, in such
  a way  a forward sub-process may essentially go unidimensionally, like in a
  ballistic motion, whereas its reverse counterpart would explore the 
whole three-dimensional
  domain. This situation typically appears in the case of matter 
desorption, in which detachment
  occurs part by part  from a "reactive" surface spot.

Some other confirmation of (mean) harmonicity,   its close relation 
to the golden rule,
  and to the Fibonacci sequencing (characterizing well the fractality 
of "diffusive" microstructures),
   is hidden in the (macro)ion (or, cluster \cite{chemph}) diffusion 
coefficient $D_\alpha $, in our model,
   Eq. (\ref{Dalfa}), which is also included in the free energy $\Phi 
$. The label, or the lower index, is
   simply $\alpha $, which for $d = 1$ results in
$\alpha = 0 \equiv 0/1$, for $d = 2$ gives
$\alpha = 1/2$, whereas for $d = 3$ offers
$\alpha = 2/3$. The values of the diffusion coefficient (Eq. 
(\ref{Dalfa})) are: $D_0\equiv {D/1}$,
  $D_{1/2}\equiv {D/2}$ as well as to $D_{2/3}\equiv {D/3}$, 
respectively. They correspond to the first five-number
  Fibonacci sequence, composed of the numerators and denominators of 
$\alpha$-s, like 0,1,1,2,3, and obe
   $a_{n+2} = a_{n+1} + a_{n}$, for the three subsequent Fibonacci 
numbers $a_{n}$, $a_{n+1}$ and $a_{n+2}$.
   If so, one can provide the following two recursive formulae
\begin{eqnarray}       \label{Fib1}
{{\alpha }^{(d-1)}} = {{a_{d-1}}\over a_d}, \qquad d = 1 ,
\end{eqnarray}

and
\begin{eqnarray}       \label{Fib23}
{{\alpha }^{(d-1)}} = {{a_{d}}\over a_{d+1}}, \qquad d = 2,3 ,
\end{eqnarray}
where $a_0 = 0$, $a_1 = 1$, $a_2 = 1$, $a_3 = 2$, $a_4 = 3$ are the 
first five Fibonacci numbers.
Since the analogy with gelling systems seems evident \cite{chemph}, 
this cannot be taken entirely as a surprise.
The bonding in gels   clearly goes as a branching process, being (as 
in the case of ultrametric space)
  quite naturally described geometrically in terms of Fibonacci 
numbers, thereby involving   the notion of fractality  \cite{weitz}.

When finishing this section, let us note that both the characteristic 
chaotic measures, cf. Eq. (\ref{LPsp2})
for example, have their random close-packing account $d+1$ involved. 
This is a landmark of randomness
  but readily appears as a space-filling action of modeled matter 
reorganisations.
  Realize that our rationale may apply just in the same vein to 
cluters-containing assemblies,
  evolving in a $d$--dimensional space, where a cluster is 
characterized by its fractal dimension $0 < {d_F} < d$, cf. 
\cite{chemph,chemphys}.


\section{Chaos in an infinite-dimensional agglomerating  and/or 
aggregating  system}
\label{sec:6}
Consider the case $\lim_{d\to\infty}$. A corresponding chaotic 
measure for the late-stage growing event
  in the agglomeration of matter, very reminiscent of nonequilibrium 
correlational entropy measure \cite{evans}, reads
\begin{eqnarray}       \label{LPspi}
{{\nu _{sp}}^{(\infty )}} = \lim_{d\to\infty} {{\Big({ln{[{\sigma }^2 
(t)]}\over {ln(t)}}\Big)}_{{for}\quad t>>1}} ,
\end{eqnarray}
whereas its counterpart for the relaxation is given by an analogous 
formula, namely that
\begin{eqnarray}       \label{LPreli}
{{\mu _{sp}}^{(\infty )}} = \lim_{d\to\infty} {{\Big({- ln{[{\sigma 
_m} (t)]}\over {ln(t)}}\Big)}_{{for}\quad t>>1}} ,
\end{eqnarray}
holds. They are consistent formally with the so-called correlational 
entropy (Kolmogorov--type) measure, defined in  \cite{schuster} and
  follow the rationale presented in \cite{rysiu}, in which some 
measures of chaos in dynamical systems
  described by partial differential equations have been discussed.
  For "thermostatic" systems out of equilibrium one has to speak of 
the so-called generalized fractal dimension formalism,
  first introduced by Grassberger and Proccacia, see \cite{schuster}, 
and refs. therein.

The most attractive reason for introducing such measures arises
from the fact that if one evaluates both ${{\nu _{sp}}^{(\infty )}}$ 
and ${{\mu _{sp}}^{(\infty )}}$,
one unambigiously gets for the aggregation
\begin{eqnarray}       \label{LPmeas0}
{{\nu _{sp}}^{(\infty )}} = {{\mu _{sp}}^{(\infty )}} = 0 ,
\end{eqnarray}
whereas for the close-packed agglomeration one provides
\begin{eqnarray}       \label{LPmeas1}
{{\nu _{sp}}^{(\infty )}} = 1 ,
\end{eqnarray}
and
\begin{eqnarray}       \label{LPmeasinf}
{{\mu _{sp}}^{(\infty )}} = \infty .
\end{eqnarray}
Thus, for both   cases, Eq. (\ref{LPmeas1}) and Eq. (\ref{LPmeasinf}),
one arrives at a chaotic behavior in the nonequilibrium system 
\cite{evans,schuster} of
a densely-packed agglomerate. This is not the case of the aggregation 
for which the common measure
  of its chaotic character is zero, cf. Eq. (\ref{LPmeas0}).

Thus, proceeding consistently with the approach offered in 
\cite{rysiu} we may conclude that the
late-time $aggregation$ process develops in an ordered way. The case 
$d=2$ appears to be the most
efficient since the harmonic-mean rule (\ref{HR}) is applied; for it the 
nonequilibrium character of the random
process should be emphasized \cite{udo}. It is intriguing to realize 
that the system property called
the harmonicity  throughout is very much related to its nonequilibrium entropic 
or chaotic characteristic(s).
%
\section{Concluding address}
\label{sec:7}
Based on the above, we are allowed to state the following: \\
(i) in matter-agglomerating systems chaos is revealed as a complex
  spatio-temporal and temperature-dependent phenomenon; \\
(ii) nonequilibrium chaotic measures of any late-stage matter agglomeration
modeled can be proposed relying upon the nonequilibrium 
Kolmogorov-type entropy measure,
which makes a reliable (harmonic) quantification of the tempo of the 
process; \\
(iii) coupling late-stage matter agglomeration with relaxation of 
assisting elastic fields
via an Onsager-type \cite{lars}, or, in the parlance of physical 
metallurgy, H-P-G conjecture \cite{HP},
  leads to several characteristic sub-effects (Bethe-lattice 
generator, first-five Fibonaci-number signatures, random close-packing $d+1$--criterion \cite{zallen}, etc.)
   having their rationale in fundamental properties of the entropic or 
harmonic-mean character of the
   phenomenon\footnote{As can be for example observed in clays made 
of an inorganic material known as laponite \cite{olivier}}; \\
(iv) as for the formal point of view: The presented mesoscopic 
system, Section 2, serving to describe
the matter aggregation can be derived rigorously based on the Gibbs 
entropy production equation \cite{chemph,vilar,chemphys},
and \\ (v) its chaotic signatures can be inferred as presented in 
Sections 3--6,
supported somehow by the ideas contained in \cite{rysiu}; at this point, a general task  remains to be done as to connect the type of chaos with the entropy-based scheme \cite{vilar,chemph} 
used to derive the equations of F-P-K \cite{FPK}, or diffusion, types \cite{lutz,chemphys,physica}, and how far the 
proposed measures of chaos (also, the ones used in the present review) are reminiscent of those used conventionally in nonlinear science \cite{dorfman,schuster,evans,lutz,udo}? Perhaps, the Edwards' entropy measures for slowly moving grains, evolving (bio)polymer- or colloid-type matrices and compacted powders could also contribute to solve the problem \cite{barrat}. 

\section{Acknowledgement}

One of us (A.G.) dedicates this study to Prof. Peter Laggner, 
\"O.A.W., Graz, Austria, and Prof. Gerard Czajkowski, U.T.A. 
Bydgoszcz,
  Poland, in the year of their  60$^{th}$ anniversary.
This work is done under 2P03B 03225 (2003-2006) by A.G. Part of M.A. 
works is in the framework of an Action de Recherches
Concert\'ee Program of the University of Li$\grave e$ge (ARC 02/07-293).
Thanks go to Prof. Peter Richmond (Dublin) for deciding to support 
the participation of A.G. in the meeting "Verhulst 200 on Chaos",
  16--18 September 2004,  Royal Military Academy, Brussels, Belgium 
from the funds of COST P10.
   Last but not least we are thankful to Prof. Jerzy \L uczka 
(Katowice) for drawing our attention to Ref. \cite{bimberg} and Prof. Miguel Rub\'{\i} (Barcelona) for useful comments on the manuscript.




\begin{thebibliography}{99.}
%
%
%
\bibitem{ma2} M.~Ausloos, Solid State Commun. \textbf{59}, 401 (1986)

\bibitem{evans} D.~J.~Evans, G.~P.~Morriss: \textit{Statistical Mechanics of NonEquilibrium Liquids} (Academic Press, London 1990)

\bibitem{HP} P.~O.~Hall: Proc. Roy. Soc. B \textbf{64}, 747 (1951); N.~J.~Petch: Phil. Mag. \textbf{1}, 186 (1956); A.~A.~Griffith: Phil. Trans. Roy. Soc. (London) A \textbf{221}, 163 (1920) 

\bibitem{lars} L.~Onsager: Phys. Rev. \textbf{37}, 405 (1931) 

\bibitem{weitz} S.~Manley et al.: Phys. Rev. Lett. \textbf{93}, 108302-1 (2004) 

\bibitem{rysiu} R.~Rudnicki: Math. Meth. Appl. Sci. \textbf{27}, 723 (2004)

\bibitem{kaye} B.~H.~Kaye: \textit{A Random Walk Through Fractal Dimensions} (VCH Verlagsgesellschaft, Weinheim 1989) 

\bibitem{physica} M.~Niemiec, A.~Gadomski, J.~\L uczka, L.~Schimansky--Geier: Physica A \textbf{248}, 365 (1998)  

\bibitem{chemphys} A.~Gadomski, J.~M.~Rub\'{\i}: Chem. Phys. \textbf{293},  169 (2003) 

\bibitem{vilar} J.~M.~G.~Vilar, J.~M.~Rub\'{\i}: Proc. Natl. Acad. USA \textbf{98}, 11081 (2001) 

\bibitem{FPK} G.~M.~Zaslavsky:  Phys. Rep. \textbf{371}, 461 (2002)

\bibitem{pichor} K.~Pich\'or, R. Rudnicki: J. Math. Anal. Appl. \textbf{215}, 56 (1997) 

\bibitem{vicsek} I.~Der\'enyi: Stochastic processes. In: \textit{Fluctuations and Scaling in Biology} chap. 2.2, ed. by T. Vicsek (Oxford University Press, Oxford 2001) pp. 27--31 

\bibitem{dorfman} J.~R.~Dorfman: \textit{An Introduction to Chaos in Nonequilibrium Statistical Mechanics} (WN PWN Warsaw 2001, \textit{in Polish}) chap. 6.2 

\bibitem{ivan} J.~M.~Rub\'{\i}, I.~Santamar\'{\i}a--Holek, A.~P{\'e}rez--Madrid: J. Phys. C \textbf{16}, S2047 (2004) 

\bibitem{bimberg} D.~E.~Jesson, T.~P.~Munt, V.~A.~Shchukin, D.~Bimberg: Phys. Rev. Lett. \textbf{92}, 115503-1 (2004)

\bibitem{gczajk} L.~Silvestri, G.~Czajkowski, F.~Bassani,
J. Phys. Chem. Solid. \textbf{61}, 2043 (2000)

\bibitem{ma1} N.~Vandewalle, B.~Delisse, M.~Ausloos, R.~Cloots, Phil. Mag. B \textbf{78}, 397 (1998) 

\bibitem{peter} P.~Laggner, M.~Kriechbaum: Chem. Phys. Lipids \textbf{57}, 121 (1991) 

\bibitem{graz} A.~Gadomski: Phil. Mag. Lett. \textbf{70}, 335 (1994)  

\bibitem{rysiu2} A.~Gadomski, J.~\L uczka, R. Rudnicki: Physica A \textbf{325}, 284 (2003)

\bibitem{chemph} A.~Gadomski, J.~M.~Rub\'{\i}, J.~\L uczka, M.~Ausloos: Chem. Phys. \textit{in press} (2004)

\bibitem{zallen} R.~Zallen: \textit{The Physics of Amorphous Solids}, 
(John Wiley \& Sons, New York 1983) chap. 2 

\bibitem{schmelz} J.~Schmelzer, G.~R\"opke, R.~Mahnke: \textit{Aggregation Phenomena in Complex Systems} (Wiley--VCH, Weinheim 1999) 

\bibitem{henk} V.~J. Anderson, H.~N.~W.~Lekkerkerker: Nature \textbf{416}, 811 (2002) 

\bibitem{wloch} W.~Przygocki, A.~W\l ochowicz: \textit{Polymer Physics} (WN PWN, Warsaw 2001, \textit{in Polish}) chap. 4 

\bibitem{lutz} V.~S.~Anishchenko, V.~V.~Astakhov, A.~B.~Neiman, T.~E.~Vadivasova, L.~Schimansky--Geier: \textit{Nonlinear Dynamics of Chaotic and Stochastic  Systems. Tutorial and Modern Developments} (Springer-Verlag, Berlin 2002) chap. 2

\bibitem{schuster} H.~G.~Schuster: \textit{Deterministic Chaos. An Introduction} (VCH Verlagsgesellschaft, Weinheim 1988) chap. 5.2

\bibitem{pastijn} V.~A.~Kostitzin: \textit{Biologie Math\'ematique} 
(Librairie Armand Colin, Paris 1937)

\bibitem{udo} U.~Erdmann: Kollektive Bewegung. Ph.D. 
Thesis, \textit{in German}, Humboldt University of Berlin, Berlin (2004) 

\bibitem{olivier} E.~Olivier, E.~Pefferkorn: Colloid Polym. Sci. \textbf{279}, 1104 (2001) 

\bibitem{barrat} A.~Barrat, J.~Kurchan, V.~Lorreto, M.~Sellito: Phys. Rev. Lett. \textbf{85}, 5034 (2000)

\end{thebibliography}
\end{document}